\documentclass[prl,aps,a4paper,showpacs,superscriptaddress,floatfix,twocolumn]{revtex4-1}
\usepackage{amsmath}
\usepackage{amsfonts}
\usepackage{amssymb}
\usepackage{graphicx}
\usepackage{dcolumn}
\usepackage{bm}
\begin{document}

\def\Zz{\mathbb{Z}}
\def\T{\hat{T}}

\title{Bound States in the Continuum realized in
the one-dimensional Two-Particle Hubbard Model with an Impurity}
\author{J.~M. Zhang,$^1$ Daniel Braak,$^2$ and Marcus Kollar}
\affiliation{Theoretical Physics III, $^2$Experimental Physics VI,
Center for Electronic Correlations and Magnetism,
University of Augsburg, 86135 Augsburg, Germany}

\begin{abstract}
We report a bound state
of the one-dimensional two-particle (Bose or Fermion) Hubbard model with an impurity potential. This state has the Bethe-ansatz form, although the model is nonintegrable. Moreover, for a wide region in parameter space, its energy is located in the  continuum band. A remarkable advantage of this state with respect to similar states in other systems is the simple analytical form of the wave function and eigenvalue. This state can be tuned in and out of the continuum continuously.
\end{abstract}

\pacs{71.10.Fd, 03.65.Ge, 03.75.-b}
\maketitle

A widespread preconception in quantum mechanics is that a normalizable (bound) state cannot be degenerate in energy with a non-normalizable (extended) state. Generically, this is indeed the case: As argued by Mott in favor of the existence of sharp mobility edges \cite{mott}, degeneracy between a localized  and an extended state
would be unstable against an infinitesimal perturbation  which can convert the localized state into an extended one. However, as pointed out by von Neumann and Wigner as early as in 1929 \cite{wigner}, the so-called bound state in the continuum (BIC), which is localized in space yet whose energy falls in the continuum band, does exist. Notably, von Neumann and Wigner had a simple, beautiful algorithm to construct such an exotic state together with the corresponding potential. They prescribed the state first and then sought an appropriate potential supporting it \cite{wigner, construct}. Of course, despite its simplicity in proof of concept, this strategy has the disadvantage that it is the state that
determines the potential, not vice versa. Moreover, the potential turns out to be rather complicated and not intuitive.

It was almost forty years after von Neumann and Wigner's original work that the concept of BIC surfaced again. This time, Stillinger argued that BICs can exist in a two-electron atom \cite{stillinger}, but not for physically relevant parameters. The first attempt to realize a BIC experimentally was taken by Capasso \textsl{et al}. \cite{capasso}.
More recently, BICs have been  demonstrated experimentally in photonics \cite{marinica,plotnik} using the analogy between optical systems and quantum mechanics, and they are also theoretically predicted to exist in some other systems \cite{ordonez,ladron,nakamura,molina,moiseyev,gonzalez}. In spite of
this progress, we note that BICs are fragile objects in general, and so far they have been found or engineered only in few systems. Some of them are simply protected by symmetry \cite{plotnik} or rely on  separation of variables \cite{jain,robnik}, or are constructed by the von Neumann-Wigner algorithm \cite{molina}. These BICs appear therefore somewhat artificial \cite{fano}.

In this paper, we report the discovery of a BIC in a one-dimensional two-particle (Bose or Fermion) Hubbard model with an impurity potential. Here, the model and the BIC
have several desirable features in comparison with previous models and the associated BICs. First, the model is much simpler (with short-range interaction and potential)
yet nontrival, and most importantly, not artificial, in contrast to many of the
constructions above. Second, though this model was
believed to be nonintegrable, we show that
half of the eigenstates have the Bethe-ansatz form. They are distinguished from the diffractive states by a $\Zz_2$-symmetry of the model: the Bethe (non-diffractive) states are odd under this symmetry, whereas the others
are even. Some of the former are  bound states, among them the BIC.
This leads to simple, explicit expressions of its wave function and eigenvalue, from which we see that the BIC can be tuned in and out of the continuum band by varying the model parameters.

The model we investigate consists of two identical spinless bosons (or two spin-$\frac{1}{2}$ fermions in the spin singlet space) in a one-dimensional infinite lattice. A defect is located at $x=0$, leading to a local potential $V$. The two particles interact through an on-site interaction $U$. The (orbital) wave function of the two particles is denoted as $f(x_1,x_2)$, with $-\infty \leq x_{1,2}\leq +\infty$ being integers. The exchange symmetry requires $f(x_1,x_2)=f(x_2,x_1)$. The Hamiltonian is defined by its action on a wave function
\begin{eqnarray}\label{h}
H f(x_1,x_2)&=&-\sum_{\alpha=\pm 1} [f(x_1+\alpha,x_2)+ f(x_1,x_2+\alpha) ]\nonumber \\
& &+\left[V(\delta_{x_1,0}+ \delta_{x_2,0})+ U
\delta_{x_1,x_2} \right]f(x_1,x_2).\quad
\end{eqnarray}
Here $V<0$ \cite{why} is the value of the impurity potential while $U$ is the on-site interaction between the two particles, with the hopping strength set to unity. It should be stressed that in the absence of impurity potential or particle-particle interaction, the model is solvable. However, in the presence of both impurity and interaction, it becomes nonintegrable even in the two-particle sector.

\begin{figure}[t]
\includegraphics[width= 0.3\textwidth ]{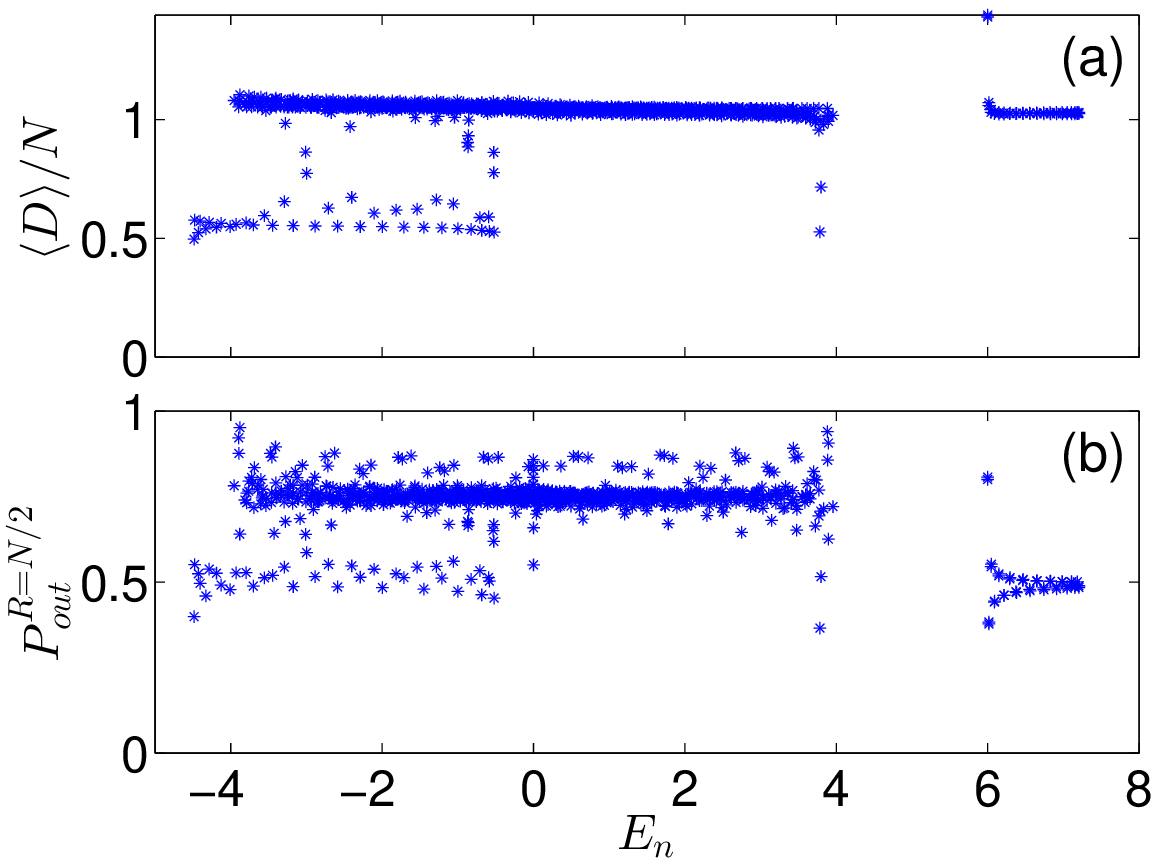}
\caption{The three continuum bands revealed by using the two quantities of $D=|x_1|+|x_2|$ and $P_{out}^{R}$. The three horizontal lines are clearly visible in each panel. The eigenstates and eigenvalues $E_n$ are solved by exact diagonalization on a $(2N+1)$-site ($N=20$) lattice and with open boundary condition. The values of the parameters are $(V,U)=(-1.5,6)$.
\label{fig1}}
\end{figure}

We aim to solve the eigenvalues and eigenstates of this system, and especially the bound states, i.e., states in which both particles are localized in the vicinity of the impurity. To demonstrate that one of the bound states lies in the continuum, we have to obtain an overview of the extended states and identify the three continuum bands associated. This classification can be verified numerically using a finite size system (see Fig.~\ref{fig1}). The first band corresponds to two particles that are neither captured by the impurity nor bound together by the interaction between them. The impurity and interaction cause phase shifts but do not contribute to the energy of the state,
 which is just the sum of the kinetic energy of both particles, and thus this band covers the interval $[-4,+4]$. For the second band, one particle is captured by the impurity but the other is not. The energy of the first particle is $-\sqrt{V^2+4}$ \cite{feynman} while that of the other
 lies between $-2$ and $2$.
This band covers therefore $[-\sqrt{V^2+4}-2,-\sqrt{V^2+4}+2]$. The third band corresponds to a delocalized molecule state \cite{repulsive}. That is, the two particles are bound together by the interaction between them and the composite moves as a whole on the lattice. The impurity causes phase shifts or local modifications on the wave function but does not contribute to the energy. This band covers $[-\sqrt{U^2+16},U]$ if $U<0$ or $[U,\sqrt{U^2+16}]$ if $U>0$ \cite{repulsive}.

The presence of these three bands can be demonstrated numerically by using some quantities which reveal the distinct nature of the extended states. The first quantity is the sum of the distance of the two particles to the defect, $D=|x_1|+|x_2|$. Suppose we choose a lattice of $2N+1$ sites with $N$ sites on each side of the defect. For the first and third bands, since the particles move through the lattice either independently or bound together, $\langle D \rangle $ should be $\sim N$. For the second band, since always one particle is localized around the defect, $\langle D \rangle $ is expected to be $\sim N/2$. The other quantity is the probability of finding at least one particle outside of a ball of radius $R$ and centered at the defect. That is, $P_{out}^R(f)=\sum_{\max\{|x_1|,|x_2|\}> R} |f(x_1,x_2)|^2$. Suppose $R=N/2$, it is easy to see that for the second and third bands, $P_{out}^R$ should be $\sim 0.5$, while for the first band it should be $ \sim 0.75$. These predictions are readily verified numerically. In Fig.~\ref{fig1}, we see how the three bands, although overlapping in energy, are separated by using $D$ and $P_{out}^R$ \cite{hunting}. Moreover, their band edges coincide with the predicted values.

\begin{figure}[t]
\includegraphics[width= 0.284\textwidth ]{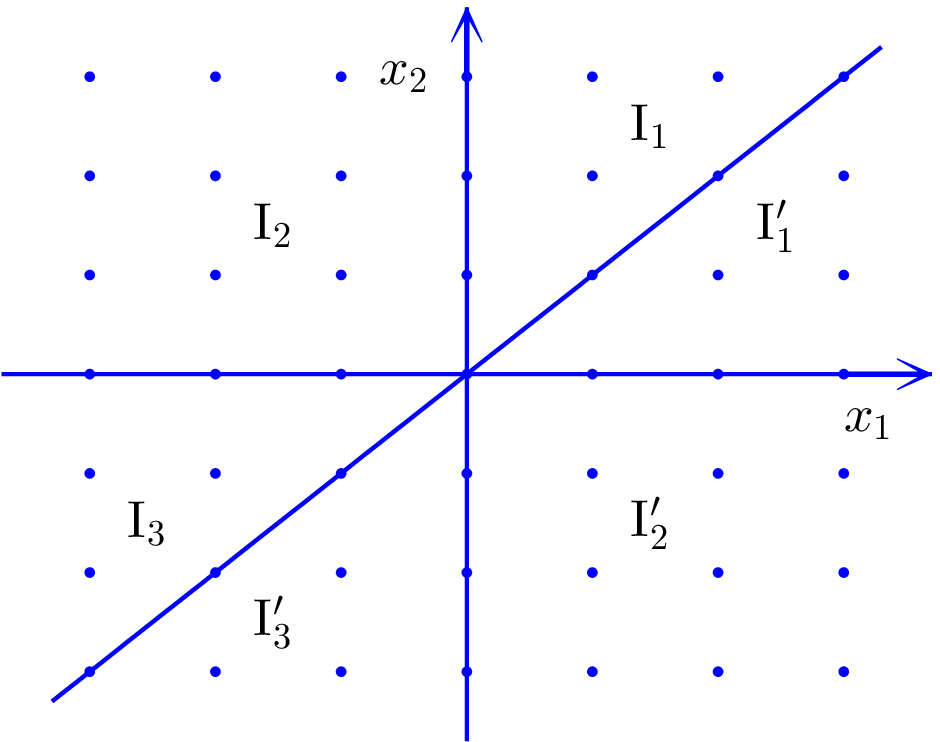}
\caption{The $x_1$-$x_2$ lattice decomposed into six regions by the $x_1$-axis, $x_2$-axis, and $x_1=x_2$ line. Note that adjacent regions share the boundary between them. Especially, the origin belongs to all the six regions.
\label{fig2}}
\end{figure}

The impurity potential and the interaction between the two particles are effective only on the three lines of $x_{1,2}=0$ and $x_1=x_2$. Away from these lines, we have free particles. This observation motivates a Bethe-type ansatz for the eigenstates, which are characterized by just two parameters, $k_1$ and $k_2$.
Specifically, in region I$_1$ ($0\leq x_1 \leq x_2 $) in Fig.~\ref{fig2}, the wave function is postulated to read
\begin{eqnarray}\label{ansatz}
f(x_1,x_2)=\sum_{\sigma_0=0,1} \sum_{\sigma_1=0,1}\sum_{\sigma_2=0,1} A_{h(\sigma_0,\sigma_1,\sigma_2)}\quad\quad \quad \quad \quad \nonumber\\
\times \exp\left[i (-)^{\sigma_1} k_1 x_{1+\sigma_0}+i (-)^{\sigma_2} k_2 x_{2-\sigma_0}\right],\quad \,
\end{eqnarray}
where $ h(\sigma_0, \sigma_1,\sigma_2)\equiv 4\sigma_0+2 \sigma_1+\sigma_2+1 $. The wave function in regions I$_2$ and I$_3$ are defined similarly but with the $A$'s replaced by $B$'s and $C$'s, respectively. The value of the wave function in the other three regions is determined by the condition $f(x_1,x_2)=f(x_2,x_1)$. In each region, we have eight different plane waves. The reason is that the interaction between the two particles can exchange their momenta, and the impurity potential can reverse the momentum of a particle. Finally, one should keep in mind that $k_{1,2}$ may be complex as below.

With the wave function in the form above, the eigenvalue equation $Hf=Ef$, with $E=-2\cos k_1 -2\cos k_2$, is satisfied away from the three lines. Now we need to fulfill it also on the three lines.
Together with the  single-valuedness of $f$ on these lines one obtains
a set of 24 homogeneous linear equations in 24 unknowns $A_j$, $B_k$, $C_l$,
$j,k,l=1\ldots 8$ \cite{supple}. The $24\times 24$ coefficient matrix, which depends on $U$, $V$, and $k_{1,2}$, needs to be, and is indeed, singular to admit nontrivial solutions. However, instead of dealing with the $24\times 24$ matrix, we employ a simplification. Note that the system is reflection invariant with respect to the impurity. Defining $[\T f](x_1,x_2)=f(-x_1,-x_2)$, we have $\T H\T=H$. Therefore, we can classify the eigenstates into even and odd ones with respect to the (parity) symmetry $\T$. It turns out that the even-parity eigenstates (especially the ground state) do not have the form of (\ref{ansatz}). However, the odd ones do \cite{supple}.

For the odd case, $f(x_1,x_2)=-f(-x_1,-x_2)$, we need
$B_i=-B_{9-i}$ and $A_i=-C_{9-i}$. It is readily verified that the former implies the latter. In the end, the linking conditions reduce to a set of homogeneous linear equations for $B_{1\leq i \leq 4}$. The $4\times 4$ coefficient matrix is always singular and has a two-dimensional nullspace \cite{supple}. The reason for the value of two is the time-reversal symmetry of the Hamiltonian.

Here some remarks are necessary. Suppose we allow both Bose and Fermi symmetry. Then Eq.~(\ref{h}) itself is invariant under both the exchange of $x_1 \leftrightarrow x_2$ and the reflection of $x_{1,2}\rightarrow -x_{1,2}$. The two symmetries divide the Hilbert space into four sectors. For the anti-symmetric (fermionic) sectors, the interaction is ineffective and we have virtually free fermions in an impurity potential. The wave functions are in the Slater form and hence also in the Bethe form. Therefore, only in one of the four sectors, i.e. the symmetric (bosonic) sector with even parity, the wave functions have not the Bethe form and are diffractive. The occurence of diffraction in a related model defined on a continuous line was shown in the classic work by McGuire \cite{mcguire}. The interesting point here is that the symmetry $\T$ leads to a decomposition of the
bosonic Hilbert space into two subspaces of which one shows no diffraction and can be therefore
considered integrable. This is confirmed by an analysis of the algebra of scattering matrices
and the associated Yang-Baxter relations \cite{next}.

We now proceed to study the odd-parity localized states. The exchange symmetry and the odd-parity condition together imply that the wave function is determined by its value in regions I$_1$ and I$_2$. After some straightforward calculation \cite{supple}, it turns out that the (unnormalized) wave function is of the form
\begin{equation}\label{wave1}
f(x_1,x_2)= e^{ik_1x_1+ i k_2 x_2}- e^{-ik_2 x_1 -i k_1 x_2},
\end{equation}
in region I$_2$, and
\begin{eqnarray}\label{wave2}
 &&f(x_1,x_2) =\frac{2V-U}{V-U}e^{ik_1x_1 +ik_2 x_{2}} -\frac{V}{V-U} e^{-ik_1x_1 + ik_2 x_{2}}\quad \nonumber\\
& &\quad \quad\quad \quad \quad\quad \quad \quad  - e^{ik_2x_1 -ik_1 x_{2}},
\end{eqnarray}
in region I$_1$. Here $k_{1,2}$ need to satisfy the equations:
\begin{equation}\label{eff}
V=z_2- z_2^{-1}, \quad V-U=z_1- z_1^{-1},
\end{equation}
and the inequalities $|z_2|<1<|z_1|<|z_2|^{-1} $, with $z_{1,2}=e^{ik_{1,2}}$. Instead of studying for what kind of $(V,U)$ pairs there are solutions of $z_{1,2}$ satisfying the inequalities, we work inversely. Because $V<0$ by assumption, we have $0<z_2<1$. Depending on the sign of $z_1$, we have two cases:

(i) $0< z_2<1<z_1<z_2^{-1}$. We get $ 0<V- U = z_1-z_1^{-1} < z_2^{-1}- z_2 =-V$. That is, $2V < U < V <0$. The energy of the wave function is $ E_{b2}=- \sqrt{V^2 +4}- \sqrt{(V-U)^2+4}$. It is easy to prove that $E_{b2}<\{-4,-\sqrt{V^2+4}-2,-\sqrt{U^2+16}\}$. Consequently, this bound state is \textit{below} all the three continuum bands and is thus \textit{not} a BIC. A notable feature of this state is that on the line $x_1+x_2=0$, $f(x_1,x_2)=0$ and if $x_1+x_2>0$, $f(x_1,x_2)<0$ while if $x_1+x_2<0$, $f(x_1,x_2)>0$. That is, the wave function has a node line $x_1+x_2=0$, and is positive on one of the two half-planes, while negative on the other. This property can be readily verified from the expressions \eqref{wave1} and \eqref{wave2}.

\begin{figure}[tb]
\includegraphics[width= 0.3\textwidth]{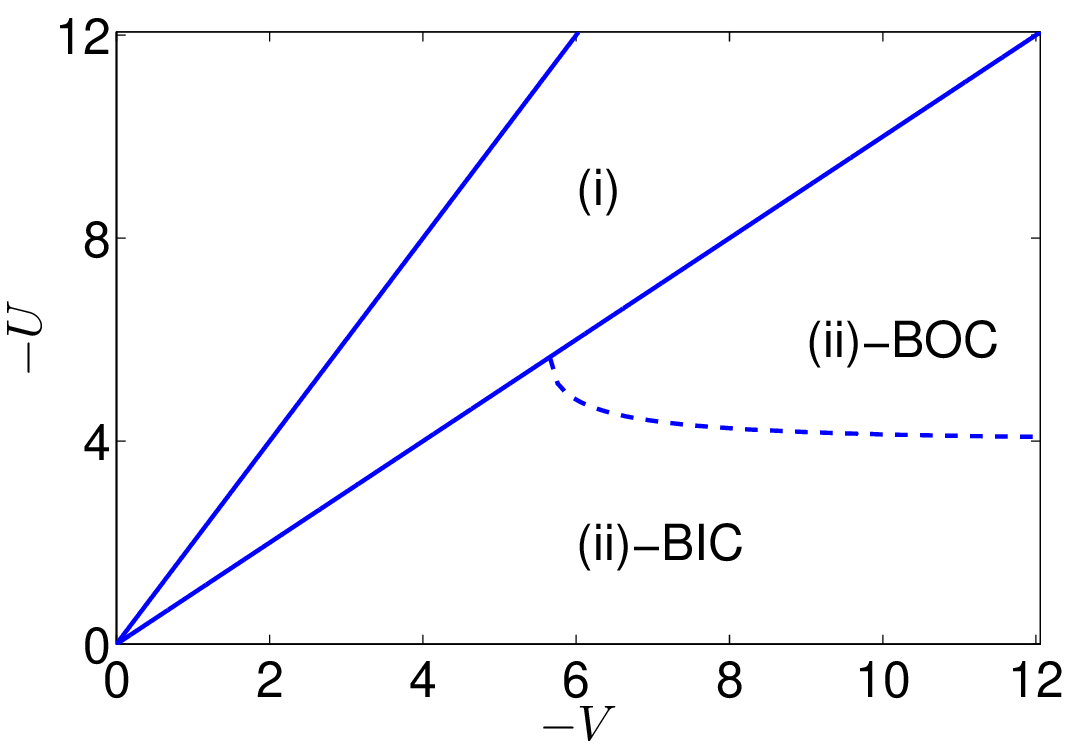}
\caption{In the region labeled with (ii)-BIC, the state with energy $E_{b1}$ [see (\ref{bic})] is a bound state embedded in the $[-4,+4]$ continuum. In the region labeled with (ii)-BOC, this state is a bound state below the $[-4,+4]$ continuum band, and outside of all the three continuum bands. The boundary between the two regions is determined by the condition $E_{b1}=-4$. In the region labeled with (i), the odd-parity bound state with energy $E_{b2}$ exists, which is below all the continuum bands.
\label{fig3}}
\end{figure}

(ii) $-z_2^{-1} <z_1 <-1<0<z_2<1$. We get $
0>V- U = z_1-z_1^{-1} > -z_2^{-1}+ z_2 =V$. That is, $V < U <0$. The eigenenergy of the wave function is
\begin{equation}\label{bic}
E_{b1}=- \sqrt{V^2 +4}+ \sqrt{(V-U)^2+4}.
\end{equation}
Now, it is easy to show that $0>E_{b1}>\{ U, -\sqrt{V^2+4}+2 \}$, and thus $E_{b1}$ falls outside of (above) the second and third band. But \textit{it can fall in the $[-4,+4]$ continuum band to be an embedded eigenvalue.} For example, for $(V,U)=(-2,-0.5)$, $E_{b1}=-0.3284$, which is inside the $[-4,+4]$ continuum. The condition for this state to be a BIC is $E_{b1}>-4$. In Fig.~\ref{fig3}, we have charted the regime where this condition is fulfilled. Note that this bound state exists whenever $V<U<0$. However, only in a subset of this regime does it fall in the continuum. Its energy can be tuned continuously across the band edge.

\begin{figure}[t]
\centering
\includegraphics[width= 0.45\textwidth]{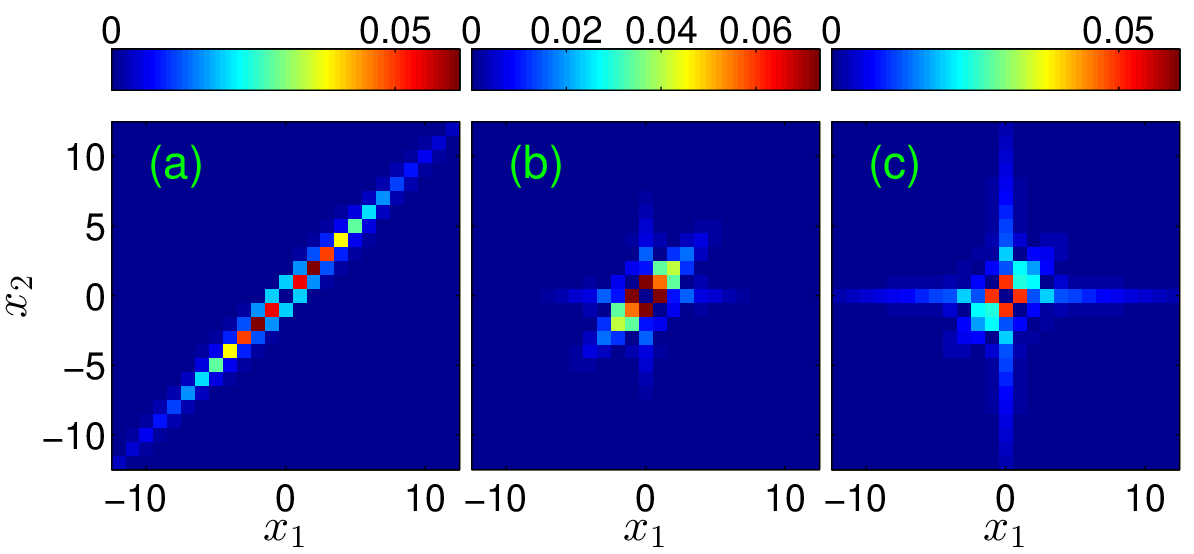}
\caption{(Color online) Images of the squared wave function of the bound state in the continuum, with (a) $(V,U)=(-2,-0.4)$,  $E_{b1}=-0.2672$, (b) $(V,U)=(-2,-1.5)$,  $E_{b1}=-0.7669$, and (c) $(V,U)=(-2,-1.8)$,  $E_{b1}=-0.8185$. These states are obtained in a finite lattice by exact diagonalization, with $100$ sites on each side of the impurity and open boundary condition. They agree with the analytic expressions in Eqs.~(\ref{wave1}) and (\ref{wave2}).
\label{fig4}}
\end{figure}

In Fig.~\ref{fig4}, we have plotted the squared wave function of the BIC for three sets of parameters. There we see that for a fixed value of $V$, the localized state gets extended along the line of $x_1=x_2$ (see Fig.~\ref{fig4}a) as $U\rightarrow 0^-$, while it gets extended along the lines of $x_1=0$ and $x_2=0$ (see Fig.~\ref{fig4}c) as $U\rightarrow V^+$.
This follows from
$|z_1z_2|\rightarrow 1$,
respectively
$|z_1|\rightarrow 1$ in the two limits. Similar behavior is displayed by the other odd-parity bound state as $U\rightarrow 2 V^+$ or $ V^-$, due to the same reason. As $U$ crosses $V$ from $V^+$ to $V^-$, the first bound state gets extended and disappears, while the second one appears starting from an infinite size. However, the two states are not continuously linked.
Actually, from (\ref{eff}) we see that $V-U$ plays the role of an effective defect potential. As it changes sign, $z_1$ changes sign and the bound state jumps discontinuously from above the second band to below it. Finally, it should be stressed that in contrast to the power-law decay of the wave function in \cite{wigner,construct}, here the wave function decays exponentially as $|x_{1,2}|$ tend to infinity, as seen in (\ref{wave1}) and (\ref{wave2}).

In conclusion, we have demonstrated the existence of a bound state in the continuum in a system of two interacting particles in an impurity potential \cite{otherb}. This discovery is actually a by-product of an investigation of the following well-motivated problem: For what value of $U>0$ will the (attractive) impurity potential no longer be able to bind the two bosons? It turns out that in this simple yet nontrivial problem  a BIC appears, which is probably the simplest nontrival BIC as far as the wave function and eigenvalue are concerned. Moreover, it should be stressed that, unlike most BICs studied before which are generally one-body BICs, here we have a two-body BIC. In the future, it would be interesting to consider also three-particle or many-particle cases to see whether ``partial integrability'' persists and many-body BICs of simple form are possible. Another direction worth pursuing is the influence of the BIC on the dynamics, e.g. the scattering properties, of the system \cite{fano}. Finally, it would be worthwhile to realize the BIC in some system experimentally. A promising candidate is guiding photonic structures \cite{marinica,plotnik,review,anderson,cold}. Note that the model (\ref{h}) can also be interpreted as a single particle hopping on a two-dimensional lattice, with potentials along three directions. Therefore, it can be simulated using a two-dimensional array of optical waveguides \cite{review, anderson}, where the hopping is realized by the evanescent coupling between neighboring waveguides, and by engineering the refractive index or geometry of the waveguides, the potentials can be realized. If the input of each waveguide is initialized according the wave function of the BIC in Eqs.~(\ref{wave1}) and (\ref{wave2}), the pattern would propagate invariantly, and thus prove the state as a BIC. Here, it should be mentioned that, since in this setting it is the propagation length that plays the role of time, it is the propagation constant that should be interpreted as the eigenenergy \cite{plotnik,review}.

We are grateful to Jiangping Hu for his interest in this problem and stimulating discussions, Michael Sekania, and Dieter Vollhardt for their helpful comments. This work was supported in part by the Deutsche Forschungsgemeinschaft through TRR 80.


\begin{thebibliography}{99}
\bibitem{why}
By the unitary transform $f(x_1,x_2)\rightarrow (-)^{x_1+x_2}f(x_1,x_2)$, we can see that the eigenvalues and eigenstates of a Hamiltonian with parameters $(V,U)$ are related to those of a Hamiltonian with parameters $(-V,-U)$ in a simple way. Therefore, in this paper we shall confine ourselves to the attractive case $V<0$.
\bibitem{hunting}
The quantities $D$ and $P_{out}^R$ can also be employed to identify bound states (e.g., the BIC): for bound states,  the values of $D$ and $P_{out}^R$ should not increase with the size of the lattice, as long as $N$ is large enough.
\bibitem{otherb}
There exist other bound states (e.g. the ground state). However, they are all of even-parity (diffractive) and located outside the continuum.
\bibitem{cold}
Another seemingly more natural system is two ultracold bosons in a one-dimensional optical lattice, where the impurity potential and on-site interaction can be readily realized and adjusted. However, since the BIC is not the ground state, it would have to be prepared dynamically, which is likely to be difficult.
\end{thebibliography}
\end{document}